\documentclass{article}

\usepackage{microtype}
\usepackage{graphicx}
\usepackage{subfigure}
\usepackage{booktabs} 

\usepackage[hyphens]{url}
\usepackage{hyperref}

\usepackage[symbol]{footmisc}

\usepackage[accepted]{icml2024}

\usepackage{amsmath}
\usepackage{amssymb}
\usepackage{mathtools}
\usepackage{amsthm}

\usepackage[capitalize,noabbrev]{cleveref}

\theoremstyle{plain}

\theoremstyle{definition}

\theoremstyle{remark}

\usepackage[textsize=tiny]{todonotes}

\usepackage{bm}

\newcommand{\dd}[1]{\,\text{d}{#1}}

\newcommand{\Ex}[1]{\text{E}\!\left[{#1}\right]}

\usepackage{color}

\usepackage{mathtools}

\usepackage{mleftright}

\newcommand{\prob}[1]{\text{P}\!\left[{#1}\right]}
\newcommand{\tprob}[1]{\tilde{\text{P}}\!\left[{#1}\right]}
\newcommand{\probc}[2]{\text{P}\mleft[ {#1} \middle| {#2} \mright] }
\newcommand{\tprobc}[2]{\tilde{\text{P}}\mleft[ {#1} \middle| {#2} \mright] }

\newcommand{\tEx}[1]{\tilde{\text{E}}\!\left[{#1}\right]}

\newcommand{\titlemanuscript}{Optimizing Watermarks for Large Language Models}
\icmltitlerunning{\titlemanuscript}

\begin{document}

\twocolumn[
\icmltitle{\titlemanuscript}



\icmlsetsymbol{equal}{*}

\begin{icmlauthorlist}
\icmlauthor{Bram Wouters}{yyy}
\end{icmlauthorlist}

\icmlaffiliation{yyy}{University of Amsterdam}

\icmlcorrespondingauthor{Bram Wouters}{b.m.wouters@uva.nl}

\icmlkeywords{Machine Learning, ICML, large language models, LLM, watermarking}

\vskip 0.3in
]



\printAffiliationsAndNotice{} 

\begin{abstract}
With the rise of large language models (LLMs) and concerns about potential misuse, watermarks for generative LLMs have recently attracted much attention. An important aspect of such watermarks is the trade-off between their identifiability and their impact on the quality of the generated text. This paper introduces a systematic approach to this trade-off in terms of a multi-objective optimization problem. For a large class of robust, efficient watermarks, the associated Pareto optimal solutions are identified and shown to outperform existing robust, efficient watermarks.
\end{abstract}

\section{Introduction}
\label{sec:introduction}
In recent years, transformer-based LLMs have proven to be remarkably powerful. Their societal impact is potentially enormous. As a consequence of their rapid rise, concerns about potential misuse have been raised. One can think of, for example, plagiarism \cite{Meyer.2023}, online propaganda \cite{Goldstein.2023}, examination in education \cite{Milano.2023}, misinformation \cite{Vincent.2022} and copyright infringement \cite{Rillig.2023}. One possible strategy to partially address these concerns is to ensure that LLM-generated text can be algorithmically distinguished from human-generated text by means of a watermark.

Initialized by \citet{Aaronson.2022} and the seminal work of \citet{Kirchenbauer.2023}, the idea of watermarking LLM-generated text has attracted much attention, in both the scientific field (see Section \ref{sec:related_work} for an overview) and the industry \cite{Bartz.2023}. Generally speaking, the process of text generation by an LLM would be adjusted in a controllable manner. Based on the generated text, a detector with knowledge of the watermarking strategy is then able to identify a text as generated by an LLM. This is usually done in the form of a hypothesis test, where the null hypothesis is that the text has been generated by a human being.

This paper largely follows the watermarking strategy of \citet{Kirchenbauer.2023}. Causal LLMs typically generate text word-by-word. Before a word is generated, the complete vocabulary of the LLM is split in two disjunct lists labelled {\it green} and {\it red}. This split is pseudo-random, where the seed is determined by the previous word(s). Green-list words are then sampled with a higher probability than the original LLM prescribes, and red-list words with a lower probability. A detector with knowledge of the pseudo-random green-red split can count the number of green-list words in a text. If this number is larger than one would expect from a text generated without knowledge of the green-red split (e.g., a human-generated text), the null hypothesis is rejected and the text is attributed to an LLM.

There are many perspectives of what a good watermark for an LLM comprises of, and even their usefulness altogether is under debate \cite{Sadasivan.2023, Jiang.2023, Zhang.2023ld2}. Roughly speaking, the quality of a watermark is assessed along four axes. A watermark must be
\begin{itemize}
\item {\bf identifiable}, meaning that a detector is able to correctly identify the generator (LLM vs.~human) of a text. 
\item {\bf stealthy}, meaning that a watermark does not noticeably change the quality of the generated text. 
\item {\bf robust} against (moderate) post-generation adjustments of the text that could obfuscate the watermark.
\item {\bf efficient} at generation and detection time, i.e., without the need for computationally costly processes.
\end{itemize}
This paper focusses on the trade-off between identifiability and stealthiness of watermarks. The probability that the hypothesis test of the detector draws the correct conclusion increases when the watermark more strongly promotes green-list tokens.
However, enforcing green-list tokens too strongly can degrade the quality of the text in unacceptable manners. We will refer to this trade-off between the quality of test and text as the {\it test-text trade-off}.

{\bf Our contribution.} For a large class of robust, efficient watermarks based on the green-red split of the vocabulary, we translate the test-text trade-off into a multi-objective optimization problem and identify the associated Pareto optimal solutions. We empirically validate the optimality of the solutions and show that they outperform existing proposals of robust, efficient watermarks \cite{Kirchenbauer.2023, Kuditipudi.2023, Wu.2023} with respect to the test-text trade-off.\footnote[1]{Code is available at \url{https://github.com/brwo/optimizing-watermarks}.} The contribution of this paper is therefore twofold. To the best of our knowledge, this is the first systematic approach to optimizing the trade-off between identifiability and stealthiness of watermarks for LLMs. Secondly, since we optimize over a large class of robust, efficient watermarks, we believe that the optimal watermarks introduced in this paper have an excellent standing with respect to the four criteria for good watermarks for LLMs.

\section{Watermarks for LLMs}
\label{sec:watermarks}
In this section a class of watermarks is defined, over which the test-text trade-off will be optimized. The class can be seen as a generalization of the watermark introduced by \citet{Kirchenbauer.2023}, based on a green-red split of the vocabulary.

In the context of LLMs, a text is typically represented as a sequence of tokens. Consider the sequence of random variables $V_1, V_2, \ldots, V_T,$ where $V_t$ corresponds to the token at position $t.$ They have support set $\mathcal{V}$ of size $N=|\mathcal{V}|,$ which is the vocabulary of the LLM. In addition, there is the prompt $V_{:1} = \left (V_0, V_{-1}, V_{-2}, \ldots \right),$ whose length is left unspecified because most modern causal LLMs do not require a fixed prompt length. The joint probability mass function (pmf) is
\begin{equation} \label{eq:joint_pdf}
\prob{V_T, V_{T-1}, \ldots V_1, V_{:1}} = \prod_{t=1}^T \probc{V_t}{V_{:t}} \times \prob{V_{:1}},
\end{equation}
where the prefix $V_{:t}$ is the subsequence of tokens prior to position $t,$ including the prompt. The causal LLM specifies $\probc{V_t}{V_{:t}},$ whereas $\prob{V_{:1}}$ represents the distribution of the text prompts under consideration. Text generation occurs token-by-token through sampling from the conditional pmf $\probc{V_t}{V_{:t}}.$

Before defining the watermark, we introduce a function $g_\gamma: \mathcal{V}^* \to \Theta,$ where $\mathcal{V}^*$ is the space of all possible prefixes of arbitrary length and $\Theta$ is the space of all subsets of size $\lfloor \gamma N \rfloor$ of the vocabulary $\mathcal{V}.$ The hyperparameter $\gamma \in (0,1)$ is fixed. Given a prefix $v_{:t} \in \mathcal{V}^*,$ the set $\mathcal{G}_t = g_\gamma(v_{:t})$ contains the so-called green-list tokens. The tokens in the complement $\mathcal{V} \setminus \mathcal{G}_t$ are the red-list tokens. This partitioning of the vocabulary is pseudo-random, with a  seed determined by a hash of $v_{:t}$ and a key. The detector of the watermark has the key and is therefore able to reconstruct the list of green tokens for each position $t$ in the sequence.

This paper considers a large class of watermarks, defined by the conditional probability distribution
\begin{equation} \label{eq:watermark}
\tprobc{V_t}{V_{:t}} = \probc{V_t}{V_{:t}} \times  \left\{
\begin{matrix}
1 + \frac{\Delta(p_t, \mathcal{G}_t)}{\Gamma_t} & \text{if } V_t \in \mathcal{G}_t, \vspace{0.2cm} \\ 
1 - \frac{\Delta(p_t, \mathcal{G}_t)}{1-\Gamma_t} & \text{if } V_t \notin \mathcal{G}_t,
\end{matrix}
\right.
\end{equation}
where $p_t$ is the function $p_t(v) = \probc{V_t=v}{V_{:t}}$ for $v \in \mathcal{V},$ representing the conditional pmf of the LLM at position $t,$ and $\Gamma_t = \probc{V_t \in \mathcal{G}_t}{V_{:t}}$ is the conditional probability that token $t$ is a green-list token. A watermark is specified by a so-called {\it shift function} $\Delta : \Xi \times \Theta \to [0,1],$ where $\Xi$ is the space of all pmfs over the vocabulary $\mathcal{V}.$ By demanding that $\Delta({p}_t, \mathcal{G}_t) \geq 0,$ the shift function increases green-list probabilities and decreases red-list probabilities. To be concrete, $\tprobc{V_t \in \mathcal{G}_t}{V_{:t}} = \Gamma_t + \Delta(p_t, \mathcal{G}_t),$ i.e., the shift function is the increase due to the watermark of the conditional probability that a token is on the green list. For $\tprobc{V_t}{V_{:t}}$ to be a valid conditional pmf, we also must demand that $\Delta(p_t, \mathcal{G}_t) \leq 1 - \Gamma_t,$ where it is important to realize that $\Gamma_t, p_t$ and $\mathcal{G}_t$ are all functions of the prefix $V_{:t}.$ A watermarked LLM generates text by sampling from $\tprobc{V_t}{V_{:t}}.$

We believe that the watermarks under consideration here, defined by Equation \eqref{eq:watermark}, have two important conceptual benefits in terms of their simplicity. First of all, the shift function $\Delta(\cdot,\cdot)$ is not a function of $V_t$ and therefore the watermark does not alter the relative probabilities among green-list tokens, i.e., $\tprobc{V_t}{V_{:t}, V_t \in \mathcal{G}_t} = \probc{V_t}{V_{:t} , V_t \in \mathcal{G}_t}.$ In other words, the watermark rescales the probabilities of all green-list tokens by the same factor, and vice versa for red-list tokens. In this sense, the change to the conditional probability distribution due to the watermarks is minimal. Secondly, the alteration of the conditional probabilities of the LLM due to the watermark is solely determined by the conditional probabilities themselves. This must be contrasted by several recent proposals for watermarks that also use a green-red split of the vocabulary, whose watermarking strategy aims to address the test-text trade-off by means of an external model \cite{Fang.2023, Li.2023, Chen.2023} and/or a word similarity score defined by a metric on the embedding space of word vectors \cite{Fu.2023, Chen.2023}.

\subsection{The KGW Watermark}
Arguably the simplest example of the class of watermarks defined by Equation \eqref{eq:watermark} is the so-called {\it hard} watermark, $\Delta_{\text{HARD}}(p_t, \mathcal{G}_t) = 1 - \Gamma_t,$ implying that $\tprobc{V_t \in \mathcal{G}_t}{V_{:t}} = 1.$ Green-list tokens are generated with probability one. This watermark is maximally strong, but at the same time impacts the text quality in an unacceptable manner \cite{Kirchenbauer.2023}. 

This is mitigated by the introduction of what we call the KGW watermark, after the first three author names of \citet{Kirchenbauer.2023}. Green-list logits are shifted by a watermark parameters $\delta \geq 0,$ while red-list logits are left unaltered,
\[
\tilde{\text{P}}_{\text{KGW}}\mleft[ V_t \middle| V_{:t} \mright]
= \frac{\exp [\ell(V_t|V_{:t}) + \delta \,\text{I}_{\mathcal{G}_t}(V_t)]}{\sum_{V_t' \in \mathcal{V}} \exp [\ell(V_t'|V_{:t}) + \delta \,\text{I}_{\mathcal{G}_t}(V_t')] } ,
\]
where $\ell(V_t|V_{:t})$ are the logits of the original LLM and $\text{I}_{\mathcal{G}_t}(\cdot)$ is an indicator function. As already mentioned, the KGW watermark is a member of the class of watermarks defined by Equation~\eqref{eq:watermark}, with
\[
\Delta_{\text{KGW}}(p_t, \mathcal{G}_t) = \frac{\Gamma_t(1-\Gamma_t) (e^\delta-1)}{1 - \Gamma_t + \Gamma_t e^\delta}.
\] The parameter $\delta$ controls the test-text trade-off, where a large $\delta$ corresponds to a high watermark identifiability.

The logic behind the KGW watermark is that green-list tokens are only substantially favored if this does not hurt text quality. Selecting a green-list token would hurt text quality if all meaningfully probable tokens are on the red list. But these tokens have much higher logits and a moderate shift $\delta$ of green-list logits will not change that. We emphasize that this particular choice of watermark is based on a heuristic argument regarding the test-text trade-off, rather than an optimization objective.

The KGW watermark is generally considered to be robust \cite{Shi.2023, Kirchenbauer.2023ba3, Piet.2023} and efficient \cite{Wu.2023}. Other watermarks defined by Equation \eqref{eq:watermark} only differ from the KGW watermark through the shift function, which does not impact robustness. Instead, robustness of watermarks based on a green-red split is typically determined by the choice of the green-list generator $g_\gamma$ \cite{Liu.2023ces}. And unless the shift function is computationally expensive, which will not be the case in the applications discussed in this paper, all watermarks defined by Equation \eqref{eq:watermark} have a comparable efficiency. We therefore conclude that the watermarks introduced in this paper can be considered robust and efficient.

\section{Optimizing Watermarks}
\label{sec:optimized_watermarks}
Optimization of the test-text trade-off requires a precise definition of both test and text quality. A simple criterion for a good test is a high number of generated green-list tokens, compared to the baseline of the non-watermarked LLM. Let $N_g$ be the number of green-list tokens in the sequence $V_1, V_2, \ldots, V_T.$ The expected number of green-list tokens shifts, as a consequence of the watermark, by $\Delta N_g = \tEx{N_g} - \Ex{N_g},$ where $\Ex{\cdot}$ is the expectation with respect to the joint pmf of Equation~\eqref{eq:joint_pdf} and $\tEx{\cdot}$ is the watermarked counterpart. It follows that
\begin{equation} \label{eq:shift_Ng}
\Delta N_g = \sum_{t=1}^T \tEx{\Delta(p_t, \mathcal{G}_t)},
\end{equation}
as the shift function is the increase (due to the watermark) of the probability that the token is a green-list token.

One common measure for text quality of an LLM is the perplexity, which is the exponential of the negative (normalized) log-likelihood. We consider the log-perplexity
\[
\log \text{PPL} = - \frac{1}{T} \sum_{t=1}^T \log \probc{V_t}{V_{:t}},
\]
and note that a high text quality corresponds to a low log-perplexity. The shift in expected log-perplexity due to the watermark, $\Delta \! \log \text{PPL} = \tEx{\log \text{PPL}} - \Ex{\log \text{PPL}},$ is given by
\begin{equation} \label{eq:shift_logPPL}
\Delta\! \log \text{PPL} = \frac{1}{T} \sum_{t=1}^T \tEx{\Delta(p_t, \mathcal{G}_t) B(p_t, \mathcal{G}_t)},
\end{equation}
where
\[
B(p_t, \mathcal{G}_t) = \sum_{v \in \mathcal{V}} \frac{\Gamma_t - \text{I}_{\mathcal{G}_t}(v)}{\Gamma_t (1 - \Gamma_t)} p_t(v) \log p_t(v). 
\]
This quantity $B(p_t, \mathcal{G}_t)$ is the expected rate of change of the log-perplexity, given the prefix $V_{:t},$ due to a shift in the conditional probability that the token is a green-list token. Roughly speaking, $B(p_t, \mathcal{G}_t)$ is large when there are no or few green-list tokens with a (relatively) large probability. It should be interpreted as the expected damage that promoting green-list tokens has on the text quality. Equations~\eqref{eq:shift_Ng} and \eqref{eq:shift_logPPL} are derived under the mild assumption that expectations are unaffected by watermark-induced changes in the distribution of the prefix $V_{:t}.$ For details, see Appendix~\ref{app:optimization}.

We are now in a position to find a watermark that optimizes the text-test trade-off. Let $\Upsilon$ be the set of all shift functions $\Delta(\cdot, \cdot),$ as defined in Section~\ref{sec:watermarks}, representing the class of watermarks defined in Equation~\ref{eq:watermark}. The aim to maximize test quality and simultaneously minimize a decrease in text quality translates into the multi-objective optimization problem
\begin{equation} \label{eq:optimization_problem}
\max_{\Delta \in \Upsilon} \Delta N_g \qquad \text{and}\qquad \min_{\Delta \in \Upsilon} \Delta\! \log \text{PPL},
\end{equation}
which has Pareto optimal solutions parametrized by $\beta \geq 0,$
\begin{equation} \label{eq:optimization_solution}
\Delta_{\text{OPT}}(p_t, \mathcal{G}_t) = \left\{
\begin{matrix}
1 - \Gamma_t & \quad \text{if}\quad B(p_t, \mathcal{G}_t) \leq \beta , \\
0 & \quad \text{if}\quad B(p_t, \mathcal{G}_t) > \beta . \\
\end{matrix}
\right.
\end{equation}
We will call this the OPT watermark, or simply OPT. For a token at position $t$ there are two options. If the expected damage to the text quality is small, at most $\beta,$ then the watermark is maximally enforced by generating a green-list token with probability one. Otherwise, no watermark is imposed and the token is sampled from the original LLM. In other words, tokens that are expected to damage the text quality the least are maximally watermarked before other tokens get any watermark at all.

For an intuitive explanation and a formal proof of the Pareto optimality of the OPT watermark, see Appendix~\ref{app:proof}. We emphasize that no additional assumptions about the joint pmf in Equation \eqref{eq:joint_pdf} are needed, as the sums of Equations \eqref{eq:shift_Ng} and \eqref{eq:shift_logPPL} can be maximized/minimized token-by-token.

Note that maximally watermarking tokens for which $B(p_t, \mathcal{G}_t) < 0$ is actually favorable for the text quality, because you make the sampling more greedy towards green-list tokens and this (potentially) decreases the perplexity. In fact, in the context of LLMs without watermarks greedy sampling minimizes $\Ex{\log \text{PPL}}.$ 

\subsection{Other Optimized Watermarks} \label{sec:other_optimal_watermarks}
The choice of optimization objectives in Equation~\eqref{eq:optimization_problem} is not unique. Different objectives could lead to different optimal watermarks. When a detector performs a hypothesis test based on a sequence of size $T,$ it will typically count the number of green-list words $N_g$ and reject the null hypothesis of a human-generated text if $N_g \geq n^*.$ Here, $n^*$ is set beforehand and corresponds to a false-positive rate $\alpha^* = \prob{N_g \geq n^*}.$ This is the probability of falsely attributing a text to an LLM (type-I error). The quality of the test is then commonly quantified as the power/sensitivity $\pi_{n^*} =  \tprob{N_g \geq n^*},$ i.e., the probability of correctly identifying an LLM-generated text.

Suppose we want to maximize the power of the test for the class of watermarks of Equation~\eqref{eq:watermark}, i.e., consider the multi-objective optimization problem
\begin{equation} \label{eq:optimization_problem2}
\max_{\Delta \in \Upsilon} \pi_{n^*} \qquad \text{and}\qquad \min_{\Delta \in \Upsilon} \Delta\! \log \text{PPL}.
\end{equation}
The OPT watermark defined in Equation \eqref{eq:optimization_solution} is also Pareto optimal for this optimization problem, provided 
the following assumptions hold for all $t$ and all $t' \neq t:$
\begin{itemize}
\item[(i)] the green-red split is unbiased, $\Ex{\Gamma_t} = \gamma,$
\item[(ii)] the events $V_t \in \mathcal{G}_t$ and $V_{t'} \in \mathcal{G}_{t'},$ which can be seen as Bernoulli random variables, are independent and identically distributed for watermarked text.
\end{itemize}
The number of green-list tokens is then binomially distributed, $N_g \thicksim \text{BIN}(T, \gamma + \tEx{\Delta(p_t, \mathcal{G}_t)}),$ and this means that maximizing the power of the test is equivalent to maximizing $\Delta N_g$ (see Appendix \ref{app:measures_test_quality} for details).

It should be stressed that the required assumptions are implicitly made throughout the literature about watermarks for LLMs, whenever test quality is measured in terms of a z-score, p-value or power of a test, as this requires that $N_g$ is binomially distributed. We also emphasize that the assumptions are only about the events of a token being a green-list token, and not about the distributions of the tokens themselves. In particular, we do not assume that the tokens $V_1,\ldots, V_T$ are independent and/or identically distributed. The validity of the assumptions is further investigated in Section \ref{sec:biasedness_dependence}.

To show that our approach can lead to different optimal watermarks, consider the following possible alternative objective for text quality: minimize the expectation of
\begingroup\setlength{\jot}{-1ex}
\begin{align*}
\frac{1}{T} \sum_{t=1}^T (- & \log \probc{V_t}{V_{:t}} )^2 =  \log \text{PPL}^2  \\
& + \frac{1}{T} \sum_{t=1}^T (- \log \probc{V_t}{V_{:t}} - \log \text{PPL} )^2,
\end{align*}
\endgroup
which can be interpreted as the bias with respect to zero (squared) plus the variance. The idea behind this objective is that it seeks to reduce the overall perplexity of a sequence, but also large deviations from this overall perplexity at the level of individual tokens. When simultaneously maximizing the power of the test, the Pareto optimal watermark is now
\begin{equation} \label{eq:optimization_solution3}
\Delta_{\text{OPT}'}(p_t, \mathcal{G}_t) = \left\{
\begin{matrix}
1 - \Gamma_t & \quad \text{if}\quad B'(p_t, \mathcal{G}_t) \leq \beta' , \\
0 & \quad \text{if}\quad B'(p_t, \mathcal{G}_t) > \beta' , \\
\end{matrix}
\right.
\end{equation}
parametrized by $\beta' \geq 0,$ where
\[
B'(p_t, \mathcal{G}_t) = \sum_{v \in \mathcal{V}} \frac{\text{I}_{\mathcal{G}_t}(v) - \Gamma_t}{\Gamma_t (1 - \Gamma_t)} p_t(v) [\log p_t(v)]^2. 
\]
We will refer to this watermark as $\text{OPT}'.$

\section{Experiments}
\label{sec:experiments}

The test-text trade-off is the main focus of this paper. We start by analyzing how the OPT watermark performs in this respect against a baseline of existing proposals for efficient, robust watermarks.

In the subsequent three sections we discuss potential limitations of the general idea of using Pareto optimal watermarks, and the OPT watermark in particular. They can be read as cautionary tales. In Section \ref{sec:other_metrics} we give two examples of the fact that optimality with respect to one metric for text quality does not necessarily generalize to other metrics. We then investigate in Section \ref{sec:biasedness_dependence} the validity of the assumptions that were needed to derive the OPT watermark. Finally, in Section \ref{sec:diversity} we present a novel analysis of the effect of watermarking on text diversity.

\subsection{Comparing the Test-Text Trade-off of Watermarks}
\label{sec:experiments_test_text_trade_off}

Our experimental setup largely follows \citet{Kirchenbauer.2023}. From the C4 dataset \cite{Raffel.2020} a sample of 500 (news) articles is drawn randomly. For each text the first (at most) 200 tokens serve as prompt, while the rest is discarded. Based on a prompt, 64 sequences of length $T=30$ are generated by means of the OPT-1.3B causal LLM \cite{Zhang.2022}, or by a watermarked version thereof. With this setup both the prefix $V_{:1}$ and the watermarked sequence $V_1,\ldots,V_T$ are sampled. The fact that we restrict ourselves to $T=30$ is not a limitation. Under the assumption that the tokens $V_1, \ldots, V_T$ are identically distributed, the results obtained for sequence length $T=30$ are easily extendable to larger sequences. Following \citet{Kirchenbauer.2023}, we let the list of green tokens for position $t$ be determined by only the token at position $t-1,$ i.e., $\mathcal{G}_t = g_\gamma(v_{t-1})$. This may not be optimal in the trade-off between tampering-resistancy and invisibility \cite{Liu.2023ces}, but we consider this outside the scope of this paper. For more details about the setup, see Appendix \ref{app:experimental_details}.

\begin{figure}[ht]
\vskip 0.2in
\begin{center}
\centerline{\includegraphics[width=\columnwidth]{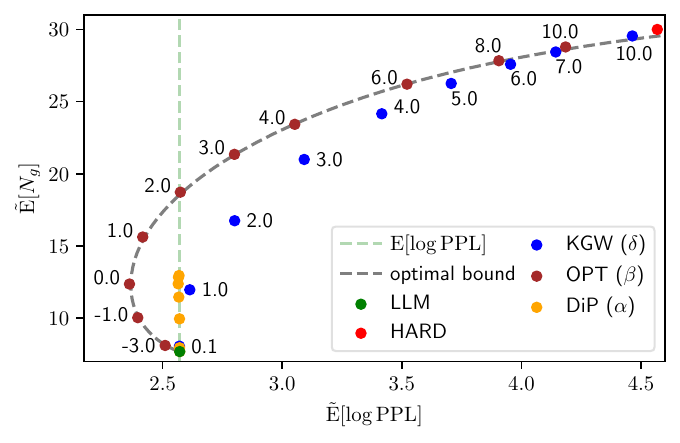}}
\caption{Test quality, measured as the expected number of green-list tokens, versus text quality, measured as the expected log-perplexity, for different watermarks. For DiP we have taken parameter values $\alpha=0.01, 0.1, 0.2, 0.3, 0.4, 0.5.$ As reference, the original language model without watermark is also included (LLM). Also shown is the Pareto optimal bound. Error bars (vertical and horizontal) are omitted, as they are never larger than the marker sizes.}
\label{fig:Ng_logPPL}
\end{center}
\vskip -0.2in
\end{figure}

\begin{figure}[ht]
\vskip 0.2in
\begin{center}
\centerline{\includegraphics[width=\columnwidth]{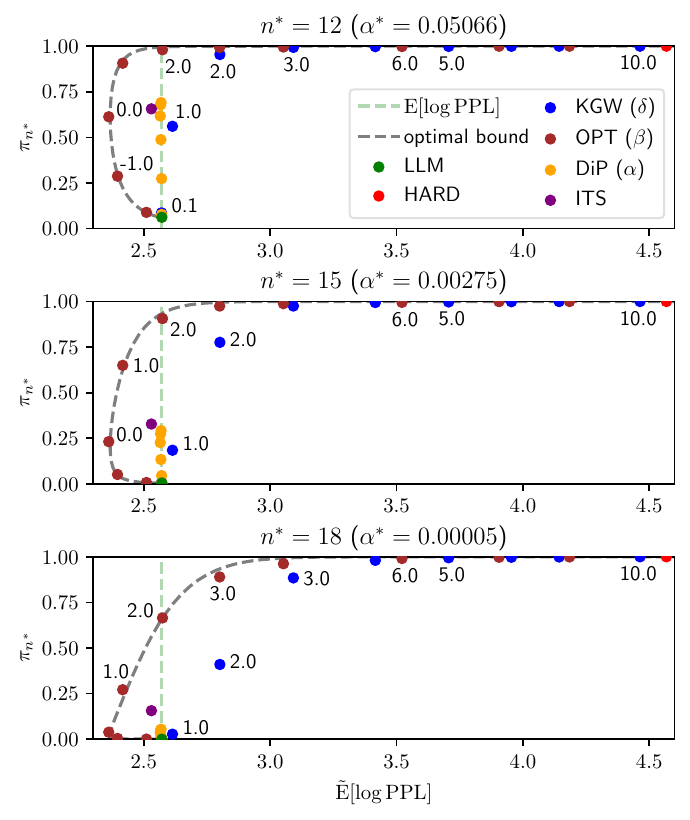}}
\caption{Similar to Figure \ref{fig:Ng_logPPL}, but now with test quality measured as the power of the test. The panels correspond to different tests ($n^*=12,15,18$) and false-positive rates ($\alpha^*$). Also included is the ITS watermark.}
\label{fig:power_logPPL}
\end{center}
\vskip -0.2in
\end{figure}

Figures \ref{fig:Ng_logPPL} and \ref{fig:power_logPPL} contain the main empirical results of this paper. Figure \ref{fig:Ng_logPPL} shows test quality, measured in terms of the expected number of green-list tokens, versus text quality, measured in terms of the expected log-perplexity. In Figure \ref{fig:power_logPPL} test quality is measured in terms of the power of the test. In addition to the hard and KGW watermarks, two more robust and (relatively) efficient watermarks are included in the baseline: the distribution-preserving watermark from \citet{Wu.2023}, with hyperparameter $\alpha,$ and the inverse-transform-sampling watermark from \citet{Kuditipudi.2023}. We will refer to them as DiP and ITS, respectively. The latter is only shown in Figure \ref{fig:power_logPPL}, as it is not based on a green-red split of the vocabulary and therefore does not have an expected number of green-list tokens. Both these watermarks have the merit of being distortion free, meaning that the watermark does not alter the sampling distribution of the LLM, when averaged over the randomness of the watermark key.

For the purpose of a fair comparison, we deliberately do not include the watermark from \citet{Kuditipudi.2023} that is based on exponential minimum sampling. It is known to incur significant computational costs at detection time and therefore cannot be considered as efficient \cite{Wu.2023, Gu.2023}. We also exclude from our analysis the two watermarks of \citet{Hu.2023}, referred to as $\delta$- and $\gamma-$reweighting. In the generation phase they are equivalent to ITS and DiP at parameter value $\alpha=0.5,$ respectively. However, their detectors make use of the weights of the LLM, which might not be available in practice. It also makes them (relatively) inefficient \cite{Kuditipudi.2023}.

Figures \ref{fig:Ng_logPPL} and \ref{fig:power_logPPL} show that OPT outperforms KGW in terms of the test-text trade-off, which was expected because of the Pareto optimality of OPT. It also outperforms DiP and ITS. This was not obvious beforehand, as the two watermarks are not in the class of watermarks defined by Equation~\eqref{eq:watermark}.

Figure \ref{fig:Ng_logPPL} also exhibits the curve of $\Ex{\Delta_\text{OPT}(p_t, \mathcal{G}_t) T}$ against $\Ex{\Delta_\text{OPT}(p_t, \mathcal{G}_t) B(p_t, \mathcal{G}_t)},$ parametrized by $\beta.$ Under the assumptions that $V_1, V_2, \ldots, V_T$ are identically distributed and that the distribution of $(p_t, \mathcal{G}_t)$ does not shift due to the presence of a watermark, this represents the Pareto optimal bound for the class of watermarks defined in Equation \eqref{eq:watermark}. The OPT watermark attains this bound, thereby validating these assumptions. We stress that this is non-trivial, as the optimal bound is computed based on properties of the original LLM only and without reference to any watermark. 
In addition, we note that the optimal bound can be used for tuning of the hyperparameter $\gamma$, which determines the fraction of green-list tokens (see Appendix \ref{app:hyperparameter_tuning} for details).

Also note that, strictly speaking, solutions with $\beta <0$ are not Pareto optimal, because $\beta=0$ is better. This means that, as long as $\beta <0,$ increasing test quality also increases text quality. In Section \ref{sec:optimized_watermarks} this was associated with the greediness of the watermarking strategy. 
Finally, we also ran the experiments for the other optimal watermark OPT$^\prime$, defined in Equation \eqref{eq:optimization_solution3}. We found that OPT and OPT$^\prime$ perform rather similarly. See Appendix \ref{app:additional_results} for the results of these experiments.


\subsection{Performance on other Text Quality Metrics}
\label{sec:other_metrics}

The performance of LLMs is commonly assessed with a variety of metrics, often depending on the specific task and type of model under consideration. Examples are perplexity, ROUGE \cite{Lin.2004} and BLEU \cite{Papineni.2002}. It is well known that these metrics measure different aspects of an LLM and are not necessarily in agreement with each other. This means that it is not guaranteed that the OPT watermark, which has been optimized for a low perplexity, also outperforms the other watermarks when text quality is measured differently. 

To illustrate this, we perform two experiments (inspired by \citet{Hu.2023}) consisting of typical tasks of LLMs: text summarization (TS) and machine translation (MT). Instead of log-perplexity, we now measure text quality in terms of ROUGE-1, which quantifies similarity between two texts in terms of the overlap of uni-grams. ROUGE scores are standard metrics to assess LLMs for summarization and translation.

For TS we use the BART-large model \cite{Liu.20204hn} and apply this to a randomly selected subset of the CNN-DM (test) dataset \cite{Hermann.2015}. For MT we use the WMT 2016 dataset and use the Multilingual BART model \cite{Liu.20204hn} to translate from English to Romanian. Both models were fine-tuned and the sample size of both experiments is 300. 

\begin{figure}[ht]
\vskip 0.2in
\begin{center}
\centerline{\includegraphics[width=\columnwidth]{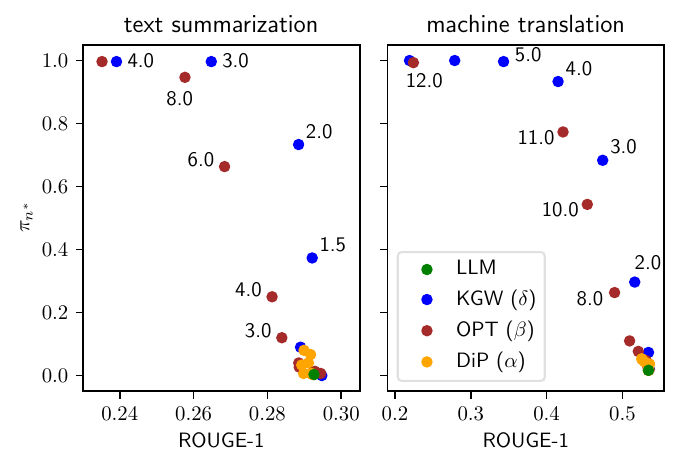}}
\caption{The power of the test versus the ROUGE-1 score, a metric for text quality, for the TS and MT tasks. For the DiP watermark the same parameter values as in Figures \ref{fig:Ng_logPPL} and \ref{fig:power_logPPL} are used. The ITS watermark has much lower ROUGE-1 scores, as it is incompatible with beam search.}
\label{fig:TS_MT}
\end{center}
\vskip -0.2in
\end{figure}

Figure \ref{fig:TS_MT} shows the test-text trade-off for both experiments and the different watermarks under consideration. Preferable is a large power of the test, corresponding to a high identifiability, and a large ROUGE-1 score close to the original LLM, corresponding to a high text quality and a stealthy watermark. We see that for TS as well as MT our OPT watermark is outperformed by the KGW watermark. This shows that watermarks, optimized for one metric, are not necessarily optimal or superior in terms of another metric. Optimized watermarks should therefore be used with caution.

Note that the ITS watermark is missing in the results. The ROUGE-1 score of ITS was very poor, as the deterministic sampling strategy is incompatible with beam search, which is standard for these types of experiments. Also note that the DiP watermark shows its distortion-free property by having a ROUGE score close to the original LLM, but performs poorly in terms of identifiability. This is in accordance with what was observed in \citet{Wu.2023}.

\subsection{Biasedness and Dependence within Watermarks}
\label{sec:biasedness_dependence}

This section investigates the validity of the assumptions that are used in the derivation of the Pareto optimal solutions to optimization problems like Equation \eqref{eq:optimization_problem2}, which involves the power of the test. The derivation uses that $N_g$ is binomially distributed, which requires two additional assumptions. The first one is unbiasedness, i.e., $\Ex{\Gamma_t} = \gamma,$ which is discussed in Appendix \ref{app:biasedness_dependence}. The second assumption is that the events that a token is a green-list token are independent and identically distributed for different tokens in the same sequence of watermarked text. The latter can be stated as that $\tEx{\Delta(p_t, \mathcal{G}_t)}$ must be the same for all $t.$ This is a mild assumption, as this is an unconditional expectation value and therefore does not depend on the prefix of token $t.$ 

\begin{figure}[ht]
\vskip 0.2in
\begin{center}
\centerline{\includegraphics[width=\columnwidth]{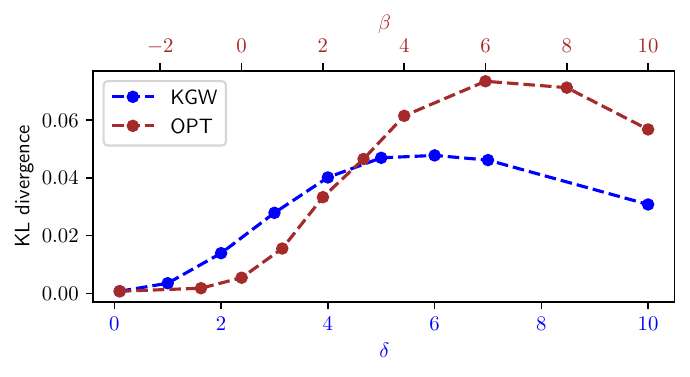}}
\caption{The Kullback-Leibler divergence between a binomial distribution for $N_g,$ based on the empirical fraction of green-list tokens, and the empirical distribution of $N_g,$ as a function of the strength ($\delta$ and $\beta$) of the respective watermarks.}
\label{fig:KL}
\end{center}
\vskip -0.2in
\end{figure}

However, it turns out that independence is violated and that the dependence becomes stronger for stronger watermarks. Figure \ref{fig:KL} shows the Kullback-Leibler divergence between an exact binomial distribution and the empirical distribution of $N_g.$ The KL-divergence increases with the strength of the watermarks, indicating a stronger (stochastic) dependence. For very strong watermarks the KL-divergence decreases again, as there is less space to deviate when almost all tokens are green-list tokens. Additional evidence of the breakdown of the binomial assumption can be found in Appendix \ref{app:biasedness_dependence}. It can be explained by the fact that not all texts allow for the same semantic freedom. Hence, the events of tokens being a green-list token are positively correlated within a sequence.

It is important to realize that Pareto optimality of the OPT watermark and the Pareto optimal bound both depend on this assumption. 
In our experiments its breakdown seems of little consequence, indicated by the fact that the OPT watermark outperforms other watermarks and that it coincides with the Pareto optimal bound in Figures \ref{fig:Ng_logPPL} and \ref{fig:power_logPPL}. But the impact might be bigger in situations with less semantic freedom, for example in code generation \cite{Wang.2023, Lee.2023}.

\subsection{The Impact of Watermarks on Text Diversity}
\label{sec:diversity}

\begin{figure}[ht]
\vskip 0.2in
\begin{center}
\centerline{\includegraphics[width=\columnwidth]{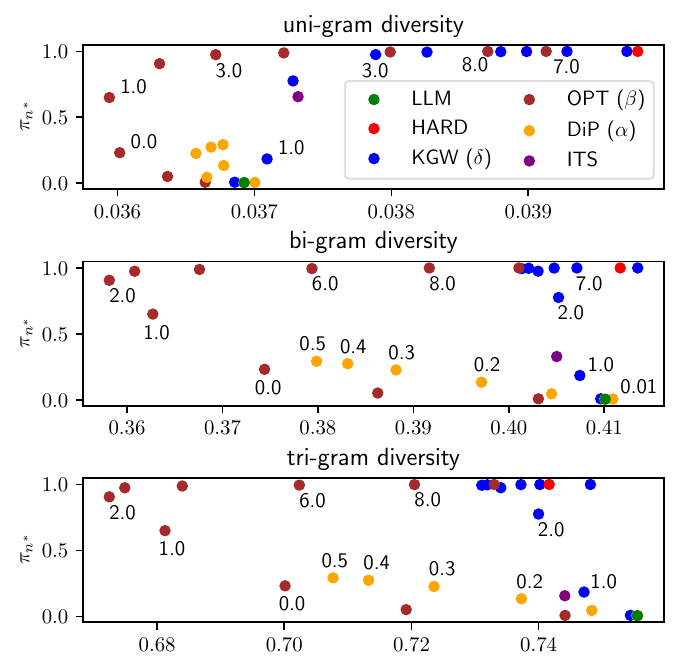}}
\caption{Uni-, bi- and tri-gram diversities for different watermarks. An n-gram diversity is the fraction of unique n-grams within the total set of n-grams of a text corpus. The results are based on the same data as Figures \ref{fig:Ng_logPPL} and \ref{fig:power_logPPL}.}
\label{fig:ngram_diversity}
\end{center}
\vskip -0.2in
\end{figure}

Watermarks like KGW and OPT are expected to decrease the n-gram diversity, which is the fraction of unique n-grams within the n-grams of a text corpus \cite{Li.2015}. The reasoning is that at position $t,$ given the prefix $V_{:t},$ the watermark (potentially) gives preference to green-list tokens, which form a randomly selected subset of the total vocabulary $\mathcal{V}.$ Also recall our interpretation of the OPT watermark with $\beta \leq 0$ as greedier than the original LLM. Figure \ref{fig:ngram_diversity} indeed shows that bi- and tri-gram diversity decrease as a consequence of the watermarks, an effect that is relatively large for the OPT watermark.

The latter is a potential disadvantage of OPT compared to other watermarks. However, n-gram diversity is not the same as semantic diversity or how humans perceive diversity. In fact, there is evidence that n-gram diversity is a poor metric for diversity judgements from humans \cite{Tevet.2020}. It requires further research to find out whether this concern about OPT watermarks is warranted. Note that, although distortion-free, DiP also shows a decrease in bi- and tri-gram diversity. It is unbiased at the level of individual tokens, but it is not invisible \cite{Liu.2023ces} when considering pairs or triplets of consecutive tokens.

Another interesting feature of Figure \ref{fig:ngram_diversity} is that, when the KGW and OPT watermarks approach the hard watermark (large $\delta$ or $\beta,$ respectively), the n-gram diversity starts to increase again. A possible explanation is that for these `harder' watermarks a more diverse set of rather unlikely green-list tokens are sampled at the expense of more standard red-list tokens. This also explains why a hard watermark has a larger uni-gram diversity than the original LLM.

\section{Related Work}
\label{sec:related_work}

Watermarking LLM-generated texts is an example of steganography, the practice of representing information (e.g., a watermark) in other information (e.g., a text). The idea of watermarking machine-learning generated text has been around for some time \cite{Venugopal.2011, Ziegler.2019} and has different realizations.

A first option is to impose a distribution shift on the LLM that is tractable for a detector, with \citet{Kirchenbauer.2023} as the most prominent example. Since then, mostly in view of mitigating between the four criteria for watermarks outlined in Section \ref{sec:introduction}, a myriad of alternatives has been proposed \cite{Zhao.2023, Wang.2023, Fang.2023, Lee.2023, Liu.2023, Fu.2023, Liu.2023ces, Chen.2023}. A notable subclass of examples, including DiP, are unbiased or distortion-free watermarks. This means that the distribution shift of the watermark is unbiased when averaged over the pseudo-random aspect of the watermark \cite{Hu.2023, Wu.2023}. 

A second option, which includes ITS, is to instill a watermark into the pseudo-random sampling from the LLM \cite{Christ.2023, Kuditipudi.2023}. This also has the advantage that it can be made distortion-free. However, examples come also with disadvantages of low robustness or inefficiency \cite{Wu.2023}.

A third option is a watermark based on an additional ML-model \cite{Abdelnabi.2021, Qiang.2023, Yoo.2023, Yang.2023, Munyer.2023}. The watermark is then imposed by making alterations to a text that is already generated by the LLM. This requires extra training and due to the extra flexibility ensuring the quality of the watermark can be difficult. Finally, one could instill a watermark into the weights of the LLM by adjusting the training procedure \cite{Li.2023ix, Gu.2023}.

Another branch of this developing field is the analysis of watermarks for LLMs. This includes the design of benchmark tasks and metrics to test the quality of watermarks \cite{Piet.2023}, some with a special focus on text quality \cite{Tang.2023, Tu.2023, Ajith.2023} or robustness \cite{Krishna.2023, Sadasivan.2023, Shi.2023, Kirchenbauer.2023ba3, Zhang.2023ld2}. 

To conclude, watermarking is not the only option to distinguish LLM-generated texts from human-generated texts. An alternative is to train a binary classifier to detect LLM-generated texts (see, e.g., \citet{Mitchell.2023}). With the rapid improvement of LLMs, this has become increasingly difficult \cite{Gambini.2022}. Another option is to let the vendor of the LLM keep a copy of all generated output and provide an API that compares a text with this database of outputs \cite{Krishna.2023}.

\section{Conclusion and Discussion}
\label{sec:conclusion}

It was posited in Section \ref{sec:introduction} that the contribution of this paper is twofold. It introduces a new watermark, OPT, corresponding to the Pareto optimal solutions of the multi-objective optimization problem into which the test-text trade-off was translated. Since the watermarks over which we optimize are generally considered robust and efficient, we believe that OPT has an excellent standing with respect to the four criteria for good watermarks for LLMs: identifiability, stealthiness, robustness and efficiency. This is notwithstanding the provisions that were made about Pareto optimal watermarks in Sections \ref{sec:other_metrics}, \ref{sec:biasedness_dependence} and \ref{sec:diversity}.

But this paper should also be read as the introduction of a systematic approach to optimizing the test-text trade-off for watermarks of LLMs. The chosen translation of the trade-off into an optimization problem is not unique, as it depends on how you quantify test and text quality. And also the class of watermarks over which we optimize, defined in Equation \eqref{eq:watermark}, is not unique. It was chosen to be based on a green-red split of the vocabulary, such that it is a generalization of the original watermark of \citet{Kirchenbauer.2023}. And its form was chosen to be so-called {\it minimal}, i.e., all green-list probabilities are rescaled by the same factor and the same holds for all red-list tokens.

It is conceivable that different choices regarding the above lead, after optimization, to more preferable watermarks. One option is to remove some of the implicit restrictions that are imposed by Equation \eqref{eq:watermark}. The shift function $\Delta(p_t, \mathcal{G}_t)$ that determines $\tprobc{V_t}{V_{:t}}$ is the same for each $V_t \in \mathcal{V},$ but this does not have to be the case. Also note that the shift function is determined by properties of token $t$ alone. One could try to make it dependent on subsequent tokens; a choice for a red token at position $t$ could enable a string of green tokens in what follows. Another possibility is to keep track of the number of green-list tokens already generated and use this to adjust the shift function.

\section*{Acknowledgements}
The author would like to thank Cees Diks and Floris Holstege for useful discussions and feedback on the manuscript.

\section*{Impact Statement}
This paper presents work whose goal is to advance the field of Machine Learning. The general topic of this paper, watermarks for large-language models, is specifically aimed at mitigating potential societal risks of Machine Learning, some of which were outlined in Section \ref{sec:introduction}. This paper aims to contribute to this.

\nocite{}

\bibliography{refs}
\bibliographystyle{icml2024}

\newpage
\appendix
\onecolumn

\section{Optimization objectives and their solutions}
\label{app:optimization}

\subsection{Measures for test quality} \label{app:measures_test_quality}
One measure for test quality is $N_g,$ the number of green-list tokens in a watermarked sequence of length $T.$ A strong test, i.e.~a test with a large identifiability, corresponds to a large number of green-list tokens. If we define binary random variables $Y_t$ such that
\begin{equation}
Y_t = \left\{
\begin{matrix}
1 &  \qquad \text{if } V_t \in \mathcal{G}_t, \\
0 &  \qquad \text{if } V_t \notin \mathcal{G}_t,
\end{matrix}
\right.
\end{equation}
then $N_g = Y_1 + Y_2 + \ldots + Y_T.$ The conditional random variables $Y_t | V_{:t}$ are Bernoulli distributed,
\begin{equation}
Y_t | V_{:t} \thicksim \text{BIN}(1, \Gamma_t)
\end{equation}
in the absence of a watermark and
\begin{equation}
Y_t | V_{:t} \thicksim \text{BIN}(1, \Gamma_t + \Delta(p_t, \mathcal{G}_t) )
\end{equation}
in the presence of a watermark. The shift, due to a watermark defined by Equation \eqref{eq:watermark}, in the expected number of green-list tokens is given by Equation \eqref{eq:shift_Ng} and can be derived as follows:
\begin{subequations}
\begin{align}
\tEx{ N_g }
& = \sum_{t=1}^T \tEx{ \tEx{ Y_t | V_{:t} } } \\
& = \sum_{t=1}^T \tEx{ \Gamma_t + \Delta(p_t, \mathcal{G}_t) } \\
& = \sum_{t=1}^T \Big\{ \tEx{ \Gamma_t} + \tEx{ \Delta(p_t, \mathcal{G}_t) } \Big\} \\
& = \sum_{t=1}^T \Big\{ \Ex{ \Gamma_t} + \tEx{ \Delta(p_t, \mathcal{G}_t) } \Big\} \label{subeq:assump} \\
& = \Ex{ N_g } +  \sum_{t=1}^T \tEx{ \Delta(p_t, \mathcal{G}_t) }
\end{align}
\end{subequations}
where in \eqref{subeq:assump} we used the assumption that expectations are unaffected by watermark-induced changes in the distribution of the prefix $V_{:t}.$ Explicitly, the assumption that $\tEx{ \Gamma_t} = \Ex{ \Gamma_t}$ means
\begin{equation} \label{eq:approximate_assumption}
\sum_{v_{:t} \in \mathcal{V}^*}  \probc{V_t \in \mathcal{G}_t}{V_{:t} = v_{:t}} \tprob{V_{:t} = v_{:t}} = \sum_{v_{:t} \in \mathcal{V}^*}  \probc{V_t \in \mathcal{G}_t}{V_{:t} = v_{:t}} \prob{V_{:t} = v_{:t}},
\end{equation}
where $\mathcal{V}^*$ is the space of all possible prefixes of arbitrary length. It should be emphasized that this is an approximate assumption, based on the idea that watermark-induced changes in the distribution of the prefix $V_{:t}$ are (approximately) averaged out in the sum over $\mathcal{V}^*$. We also emphasize that this assumption does not mean that the marginal distribution of the prefix $V_{:t}$ is unaffected by the watermark, which would be a much stronger assumption.

Another measure for test quality is the power of the test, $\pi_{n^*} =  \tprob{N_g \geq n^*}.$ If we assume that $\tEx{ \Gamma_t } = \Ex{ \Gamma_t } = \gamma,$ that the $Y_t$ are identically distributed and that they are (stochastically) independent, then
\begin{equation}
N_g \thicksim \text{BIN}(T, \gamma + \tEx{\Delta(p_t, \mathcal{G}_t)} ),
\end{equation}
because $\tEx{ Y_t } = \tEx{ \tEx{ Y_t | V_{:t} } } = \tEx{ \Gamma_t + \Delta(p_t, \mathcal{G}_t) }.$ This implies
\begin{equation}
\pi_{n^*} = \sum_{n=n^*}^T \binom{T}{n} \left( \gamma + \tEx{\Delta(p_t, \mathcal{G}_t)} \right)^n \left( 1 - \gamma - \tEx{\Delta(p_t, \mathcal{G}_t)} \right)^{T-n}.
\end{equation}
In other words, the watermark determines the power of the test only through $\tEx{\Delta(p_t, \mathcal{G}_t)}$ and does this in a monotonically increasing way. Hence, maximizing the power of the test over a class of watermarks is equivalent to maximizing $N_g.$

\subsection{Measures for text quality}
One measure for text quality is the log-perplexity, defined in Section \ref{sec:optimized_watermarks}. A large log-perplexity is interpreted as a low text quality. The shift, due to a watermark defined by Equation \eqref{eq:watermark}, in the expected log-perplexity is given by Equation \eqref{eq:shift_logPPL} and can be derived as follows:
\begin{subequations}
\begin{align}
\tEx{\log \text{PPL}}
& = - \frac{1}{T} \sum_{t=1}^T \tEx{ \log \probc{V_t}{V_{:t}} } \\
& = - \frac{1}{T} \sum_{t=1}^T \tEx{ \sum_{v \in \mathcal{V}} \tprobc{V_t=v}{V_{:t}} \log \probc{V_t=v}{V_{:t}} } \\
& = - \frac{1}{T} \sum_{t=1}^T \tEx{ \Ex{ \log \probc{V_t}{V_{:t}} \,|\, V_{:t} } } \\
& \quad - \frac{1}{T} \sum_{t=1}^T \tEx{ \sum_{v \in \mathcal{V}} \Delta(p_t, \mathcal{G}_t) \left\{ \frac{\text{I}_{\mathcal{G}_t}(v)}{\Gamma_t} - \frac{1-\text{I}_{\mathcal{G}_t}(v)}{1-\Gamma_t} \right\} \probc{V_t=v}{V_{:t}} \log \probc{V_t=v}{V_{:t}} } \\
& = - \frac{1}{T} \sum_{t=1}^T \tEx{ \Ex{ \log \probc{V_t}{V_{:t}} \,|\, V_{:t} } } + \frac{1}{T} \sum_{t=1}^T \tEx{ \Delta(p_t, \mathcal{G}_t) B (p_t, \mathcal{G}_t) } \label{subeq:defB} \\
& = - \frac{1}{T} \sum_{t=1}^T \Ex{ \log \probc{V_t}{V_{:t}} } + \frac{1}{T} \sum_{t=1}^T \tEx{ \Delta(p_t, \mathcal{G}_t) B (p_t, \mathcal{G}_t) } \label{subeq:exp_assumption} \\
& = \Ex{\log \text{PPL}} + \frac{1}{T} \sum_{t=1}^T \tEx{ \Delta(p_t, \mathcal{G}_t) B (p_t, \mathcal{G}_t) } .
\end{align}
\end{subequations}
In \eqref{subeq:defB} we used the definition of $B (p_t, \mathcal{G}_t)$, see Section \ref{sec:optimized_watermarks}. In \eqref{subeq:exp_assumption} we used a similar approximate assumption as in \eqref{subeq:assump}. See \eqref{eq:approximate_assumption} for a discussion.

\subsection{Derivation of Pareto optimal solutions} \label{app:proof}

The multi-objective optimization problem of Equation \eqref{eq:optimization_problem} can be written as
\begin{equation}
\max_{\Delta \in \Upsilon} \tEx{\Delta(p_t, \mathcal{G}_t)} \qquad \text{and}\qquad \min_{\Delta \in \Upsilon} \tEx{\Delta(p_t, \mathcal{G}_t) B(p_t, \mathcal{G}_t) },
\end{equation}
where $\Upsilon$ is the set of shift functions $\Delta : \Xi \times \Theta \to [0,1],$ where $\Xi$ is the space of all pmfs over the vocabulary $\mathcal{V},$ and $\Theta$ is the space of all subsets of size $\lfloor \gamma N \rfloor$ of the vocabulary $\mathcal{V},$ with the additional (consistency) requirement that $\Delta(p_t, \mathcal{G}_t) \leq 1 - \Gamma_t.$

The expectations $\tEx{ \cdot }$ are taken over all possible sequences $V_1, \ldots, V_T$ generated by a watermarked LLM, and all possible prompts $V_{:1}.$ However, in the definition of the optimization problem only two functions of the individual tokens play a role: $\Gamma_t$ and $B(p_t, \mathcal{G}_t).$ The optimization problem can therefore be rephrased in terms of the search for a function $h(x,y)$ of a bivariate random variable $(X,Y),$ where $X$ plays the role of $\Gamma_t$ and $Y$ plays the role of $B(p_t, \mathcal{G}_t),$ that has support $[0,1] \times \mathbb{R}:$
\begin{equation} \label{eq:optimization_problem_proof}
\max_{h(\cdot,\cdot)} \text{E}_{(X,Y)}\! \left[ h(X,Y) \right] \qquad \text{and}\qquad \min_{h(\cdot,\cdot)} \text{E}_{(X,Y)}\! \left[ h(X,Y) Y \right],
\end{equation}
where the expectations are with respect to the joint distribution of $(X,Y),$ and where the function $h(x,y)$ obeys the constraint $0 \leq h(x,y) \leq 1-x.$ We claim that
\begin{equation} \label{eq:optimization_problem_proof_solution}
h^*(x,y) = \left\{
\begin{matrix}
1 - x & \qquad \text{if } y \leq \beta, \\
0 & \qquad \text{if } y > \beta, \\
\end{matrix}
\right.
\end{equation}
which corresponds to Equation \eqref{eq:optimization_solution}, is Pareto optimal if $\beta \geq 0.$ We begin by giving an intuitive argument for this claim, which is then followed by a more formal proof.

\subsubsection*{Intuitive argument for Pareto optimality}

If $f(x,y)$ is the joint pdf of $(X,Y),$ the optimization problem of Equation \eqref{eq:optimization_problem_proof} can simply be rephrased as
\begin{equation} \label{eq:optimization_problem_proof_intuitive}
\max_{\tilde{h}(\cdot,\cdot)} \int \left[ \int \tilde{h}(x,y) \dd y \right] \dd x \qquad \text{and}\qquad \min_{\tilde{h}(\cdot,\cdot)} \int \left[ \int \tilde{h}(x,y) y \dd y \right] \dd x,
\end{equation}
where $\tilde{h}(x,y) = h(x,y) f(x,y)$ and the constraint is now $0 \leq \tilde{h}(x,y) \leq (1-x) f(x,y).$ We see that for each value of $x,$ this is a separate multi-objective optimization problem. Given a value for $x,$ we want
\begin{equation} \label{eq:optimization_problem_proof_intuitive_simplified}
\max_{\tilde{h}(x,\cdot)} \int \tilde{h}(x,y) \dd y \qquad \text{and}\qquad \min_{\tilde{h}(x,\cdot)} \int \tilde{h}(x,y) y \dd y.
\end{equation}
In words, we are looking for a function $\tilde{h}(x,\cdot)$ of only a single variable $y.$ We want its area under the curve to be maximized. At the same time, we want the volume of variable height $y$ above this area to be minimized.

Suppose we want the area under the curve to be of a certain value, how do we approach this problem? The solution is to start making the function $\tilde{h}(x,\cdot)$ maximal for the smallest possible values of $y,$ because for those values the contribution to the volume is the smallest. You continue doing this for larger and larger values of $y,$ until you have reached the desired area. This will minimize the volume. If, on the other hand, you wish the volume to be of a certain value, you take the same approach until you have reached that volume. This will maximize the area under the curve.

To make the optimization problem of Equation \eqref{eq:optimization_problem_proof_intuitive_simplified} more tangible, imagine the following analogy. You are standing at location A next to a pile of stones of variable weights. If someone asks you to bring 10 stones to location B, you choose to bring the 10 lightest stones to make your life easy. If someone asks you to bring as many stones as possible to location B, but with a total weight of 5kg. Then you again choose to bring the lightest stones until you have reached a total 5kg. In this analogy the stones are selected by the function $\tilde{h}(x,\cdot),$ which plays the role of an indicator function, whereas the weight of the individual stones is the variable $y.$

\subsubsection*{Formal proof of Pareto optimality}

We continue with a more formal proof of the Pareto optimality of the function $h^*(x,y)$ defined in Equation \eqref{eq:optimization_problem_proof_solution}. We do this by showing that a function $h^\prime(x,y)$ that obeys the same constraints and that improves the first objective in Equation \eqref{eq:optimization_problem_proof}, necessarily does worse for the second objective in Equation \eqref{eq:optimization_problem_proof}. 

Hence, assume that $\text{E}_{(X,Y)}\! \left[ h^\prime(X,Y) \right] > \text{E}_{(X,Y)}\! \left[ h^*(X,Y) \right].$ Suppose $h^\prime(x,y) \neq h^*(x,y)$ if $y \in A_1 \cup A_2,$ where $A_1 \subset (-\infty, \beta]$ and $A_2 \in (\beta, \infty).$ This means that $h^\prime(x,y) < 1 - x$ if $ y \in A_1 \subset (-\infty, \beta]$ and $h^\prime(x,y) > 0$ if $ y \in A_2 \subset (\beta, \infty).$ The initial assumption then translates into
\begin{equation}
\int_{A_1} \int_0^1 \left[ (1 - x) - h^\prime(x,y) \right] f(x,y) \dd x \dd y < \int_{A_2} \int_0^1 h^\prime(x,y) f(x,y) \dd x \dd y,
\end{equation}
where $f(x,y)$ is the joint pdf of $(X,Y).$ Note that $A_1$ can be empty. Let's now focus on the second objective, which is to minimize (assuming $\beta>0$)
\begin{subequations}
\begin{align}
\Ex{h^\prime(x,y) y}
& = \int_{-\infty}^\infty \int_0^1 h^\prime(x,y) y f(x,y) \dd x \dd y \\
& = \int_{-\infty}^\infty \int_0^1 h^*(x,y) y f(x,y) \dd x \dd y \\
& \quad + \int_{A_2} \int_0^1 h^\prime(x,y) y f(x,y) \dd x \dd y - \int_{A_1} \int_0^1 \left[ (1 - x) - h^\prime(x,y) \right] y f(x,y) \dd x \dd y \\
& > \int_{-\infty}^\infty \int_0^1 h^*(x,y) y f(x,y) \dd x \dd y \label{subeq:strict_inequality1} \\
& \quad + \beta \left[ \int_{A_2} \int_0^1 h^\prime(x,y) f(x,y) \dd x \dd y - \int_{A_1} \int_0^1 \left[ (1 - x) - h^\prime(x,y) \right] f(x,y) \dd x \dd y \right] \\
& > \int_{-\infty}^\infty \int_0^1 h^*(x,y) y f(x,y) \dd x \dd y \label{subeq:strict_inequality2} \\
& = \Ex{h^*(x,y) y}
\end{align}
\end{subequations}
where in \eqref{subeq:strict_inequality2} we used that $\text{E}_{(X,Y)}\! \left[ h^\prime(X,Y) \right] > \text{E}_{(X,Y)}\! \left[ h^*(X,Y) \right]$ and \eqref{subeq:strict_inequality1} is a strict inequality because $\text{E}_{(X,Y)}\! \left[ h^\prime(X,Y) \right] > \text{E}_{(X,Y)}\! \left[ h^*(X,Y) \right]$ is a strict inequality. Note that the second inequality is an equality in the special case $\beta=0.$ This derivation shows that the second objective is worse off. One can make a similar argument for improving on the second objective in Equation \eqref{eq:optimization_problem_proof} and showing that the first objective will then necessarily become worse.

As already explained in Appendix \eqref{app:measures_test_quality}, the multi-objective optimization problem defined by Equation \eqref{eq:optimization_problem2} has the same Pareto optimal solutions.

The watermarks OPT', defined in Equation \eqref{eq:optimization_solution3}, can be shown to be Pareto optimal in a similar fashion. The only difference is that $- \log \probc{V_t=v}{V_{:t}}$ must be replaced by $[- \log \probc{V_t=v}{V_{:t}}]^2.$

\section{Experimental details}
\label{app:experimental_details}

All experiments have been implemented by means of the Huggingface library \cite{Wolf.2020}.

\subsection{Details of the experiments described in Sections \ref{sec:experiments_test_text_trade_off}, \ref{sec:biasedness_dependence}, \ref{sec:diversity}}

Prompts are created by randomly selecting (news) texts from the C4 dataset \cite{Raffel.2020}. Only texts of at least 250 tokens are taken into account. If a text has at most 400 tokens, the final 200 tokens are removed and the remainder forms the prompt. If a text has more than 400 tokens, the first 200 tokens are used as prompt and the rest is discarded. All prompts have a length between 50 and 200 tokens.

Sampling from the LLM takes place with a temperature of 1.0. In order to generate sequences of a fixed length $T,$ the EOS token is suppressed.

\subsubsection{Estimation of the log-perplexity}
Suppose $v_1, v_2, \ldots, v_T$ is an actual realization of the random variables $V_1, V_2, \ldots, V_T.$ The log-perplexity of this sequence is
\begin{equation}
- \frac{1}{T} \sum_{t=1}^T \log \probc{V_t = v_t}{V_{:t} = v_{:t}}.
\end{equation}
In order to estimate $\tEx{ \log \text{PPL}},$ the above formula could be used. However, in order to reduce estimation noise, we used
\begin{equation}
- \frac{1}{T} \sum_{t=1}^T \sum_{v \in \mathcal{V}} \tprobc{V_t = v}{V_{:t} = v_{:t}}  \log \probc{V_t = v}{V_{:t} = v_{:t}}.
\end{equation}

\subsection{Details of the experiments described in Section \ref{sec:other_metrics}}

For the text summarization (TS) and machine translation (MT) tasks, we have used the default setup. Samples of size 300 were randomly selected from the respective test sets. We used the default sampling strategy: beam search with 4 beams.

\subsection{A note about the implementation of other watermarks}

The vanilla implementation of the ITS watermark \cite{Kuditipudi.2023} lead to a much higher log-perplexity that one would expect from a distortion-free watermark. Tokens with extremely small probabilities are sampled with unrealistically high frequencies. Even tokens with a probability of strictly zero were sometimes sampled. We believe that the underlying cause is numerical inprecision when selecting the index that makes the cumulative distribution function pass a certain threshold. To address this problem, we forbid sampling from tokens with probability below 10e-6.

Furthermore, for the ITS watermark we do not use Levenshtein distance, because their purpose is to robustify the watermark against insertions and deletions and we did not consider this.

\section{Hyperparameter tuning}
\label{app:hyperparameter_tuning}

Watermarks defined by Equation \eqref{eq:watermark} have a hyperparameter $\gamma,$ the fraction of the vocabulary $\mathcal{V}$ that is on the green list. A large $\gamma$ means that for each token relatively many green-list options are available. This makes the expected deterioration of text quality relatively small. However, when $\gamma$ is large a powerful test also requires relatively many green tokens in a sequence of length $T.$ For small $\gamma$ the trade-off is vice versa. This raises the question of an optimal value of $\gamma$ with respect to the test-text trade-off.

\begin{figure}[ht]
\vskip 0.2in
\begin{center}
\centerline{\includegraphics[width=8cm]{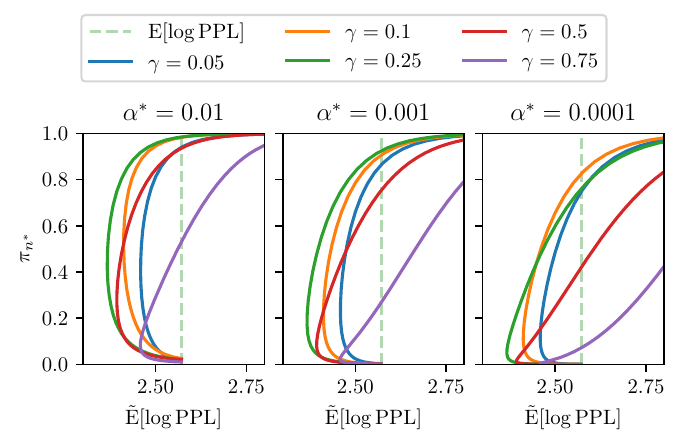}}
\caption{Pareto optimal bounds for different values of the hyperparameter $\gamma,$ for tests with different false-positive rates $\alpha^*.$ It shows that there is no universally ``best'' $\gamma.$}
\label{fig:power_logPPL_gammas}
\end{center}
\vskip -0.2in
\end{figure}

\begin{figure}[ht]
\vskip 0.2in
\begin{center}
\centerline{\includegraphics[width=8cm]{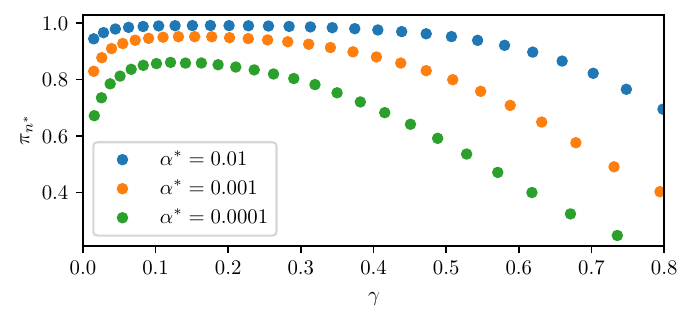}}
\caption{The power of OPT watermarks that do not affect text quality ($\,\tEx{\log \text{PPL}} = \Ex{\log \text{PPL}}\,$), as a function of the hyperparameter $\gamma,$ for tests with different false-positive rates $\alpha^*.$ The ``best'' value for $\gamma$ usually lies between 0.1 and 0.2.}
\label{fig:power_gamma_tuning}
\end{center}
\vskip -0.2in
\end{figure}

In Figure \ref{fig:power_logPPL_gammas} Pareto optimal bounds are plotted for different values of the hyperparameter $\gamma.$ It shows that there is no universally best $\gamma,$ i.e., a hyperparameter value that gives the best test-text trade-off for every possible false-positive rate $\alpha^*.$ One possible definition of a ``best'' $\gamma$ is the one that, given a false-positive rate $\alpha^*,$ maximizes the power of the OPT watermark that does not affect text quality, i.e., $\tEx{\log \text{PPL}} = \Ex{\log \text{PPL}}.$ Figure \ref{fig:power_gamma_tuning} shows that this ``best'' $\gamma$ usually lies between 0.1 and 0.2, where it should be noted that the range of ``near-best'' hyperparameter values increases with increasing $\alpha^*.$ Interestingly, \citet{Kirchenbauer.2023} found a ``best'' $\gamma$ around 0.1 for their non-optimal KGW watermark. It should be emphasized that the optimal value of $\gamma$ is not only dependent on how the test-text trade-off is defined, but is also a property of the LLM. A different LLM could give a different optimal value for $\gamma.$

\section{Additional results on biasedness and indepedence within watermarks}
\label{app:biasedness_dependence}

As mentioned in Section ~\ref{sec:other_optimal_watermarks}, the derivation of the Pareto optimal solutions to optimization problems involving the power of the test, e.g.~Equation \eqref{eq:optimization_problem2}, uses that $N_g$ is binomially distributed and this requires two additional assumptions. The first one is unbiasedness of the (conditional) probability of a token being a green-list token in the absence of a watermark, $\Ex{\Gamma_t} = \gamma.$ To test this, for 10,000 prompts from the C4 dataset a sequence of 30 tokens is generated without watermark. For each token $\Gamma_t$ is computed for a green-red split with $\gamma=0.25.$ The sample average is $\overline{\Gamma}_t = 0.2574.$ Under the null hypothesis of unbiasedness, and under the assumption of independence of the observations in the sample, this corresponds to a z-score of $16.1.$ This strongly suggests that $\Ex{\Gamma_t} \neq \gamma.$ 

\begin{figure}[ht]
\vskip 0.2in
\begin{center}
\centerline{\includegraphics[width=0.9\columnwidth]{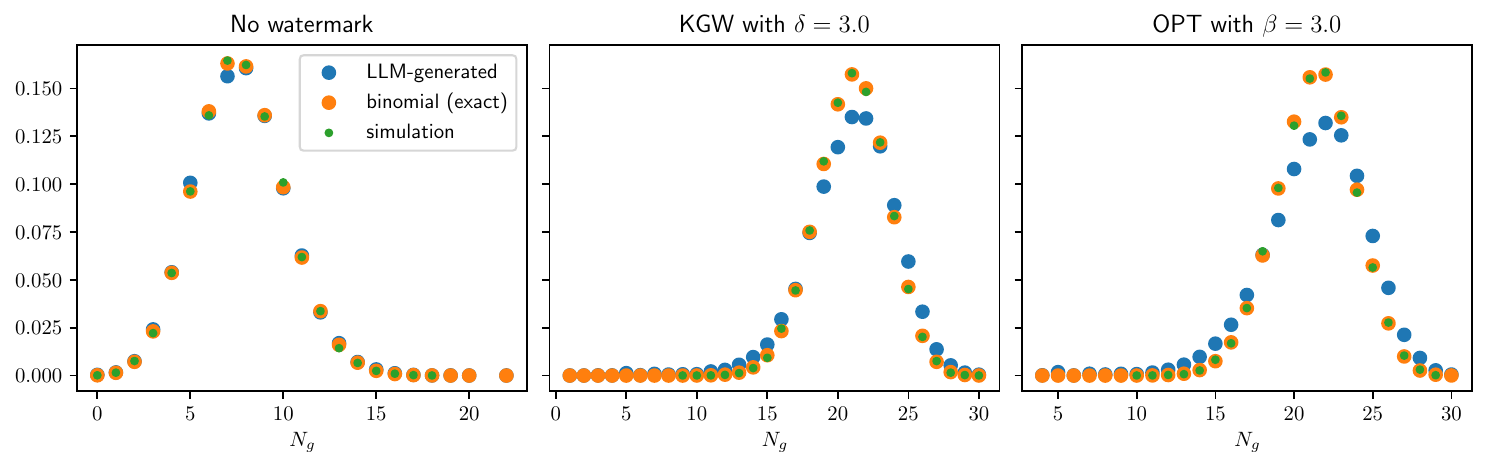}}
\caption{A comparison between the empirical distribution of $N_g,$ the number of green-list tokens in a sequence of $T=30$ tokens, and the exact binomial distribution, for text generated without watermark ({\it left panel}) and with the OPT watermark with $\beta=4.0$ ({\it middle panel}). The oracle OPT ({\it right panel}) is the OPT watermark, but with a different, randomly selected key for green-red split generation at every generation step. Because oracle OPT and OPT differ from the exact binomial distribution similarly, we conclude the discrepancy is not because of the pseudo-random green-red split. Also included is a simulation sampled from the exact binomial distribution, of the same sample size as the LLM generated data.}
\label{fig:bin_dist}
\end{center}
\vskip -0.2in
\end{figure}

This bias has its origin in the pseudo-random green-red split of the vocabulary. For each pair of subsequent tokens in a sentence, the key of the function $g_\gamma$ determines whether the second token of the pair is green or red. This is the same for each occurrence of the pair. Hence, the occurrence frequencies of all possible pairs of tokens, together with the key, determine the bias of $\Ex{\Gamma_t}$ with respect to $\gamma.$ We verified that a different generator $g_\gamma,$ based on a different key, gives a different bias. We emphasize that it is unlikely that this bias has an effect on the Pareto optimality of the solutions presented in this paper, because also in the presence of a bias the power of the test remains a monotonic function of $\Delta N_g.$

The second assumption, which was discussed in Section \ref{sec:biasedness_dependence}, is that the event that a token is a green-list token is (stochastically) independent for different tokens in the same sequence. It turns out that this is not the case, and that the dependence becomes stronger for stronger watermarks. Figure \ref{fig:bin_dist} shows additional empirical evidence that the distribution of $N_g$ for watermarked text is generally not binomial. The heavier tails indicate a positive correlation between the events that tokens from the same list are green-list tokens. This is understandable, as different sentences can have different amounts of freedom (i.e.~entropy) to insert green-list tokens.

\section{Additional results}
\label{app:additional_results}

\begin{figure}[ht]
\vskip 0.2in
\begin{center}
\centerline{\includegraphics[width=8cm]{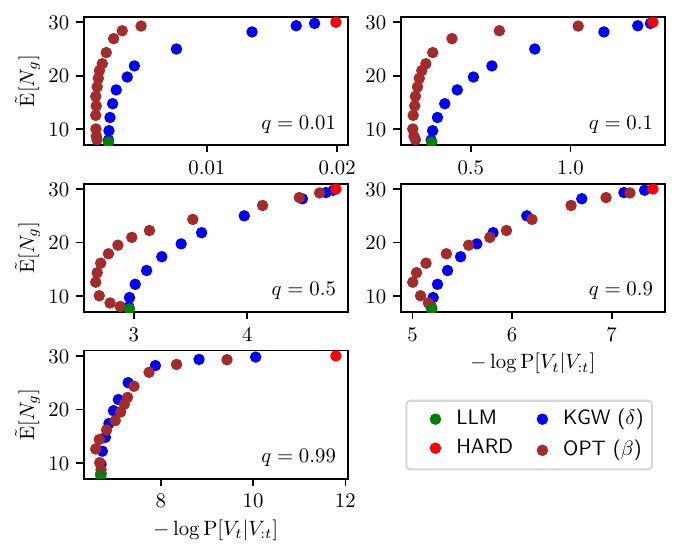}}
\caption{The $q$th percentile of $- \log \probc{V_t}{V_{:t}}$ is shown for different watermarks and $q=0.01, 0.1, 0.5, 0.9$ and $0.99.$}
\label{fig:percentiles}
\end{center}
\vskip -0.2in
\end{figure}

Additional descriptive statistics of the effect of different watermarks on text quality can be found in Figure \ref{fig:percentiles}, where the $q$th percentile of $- \log \probc{V_t}{V_{:t}}$ is plotted for different values of $q.$ The relative difference in percentiles between the KGW and OPT watermarks decreases with increasing $q$ and is virtually absent when $q=0.99.$ In other words, the OPT watermark is not better than the KGW watermark for tokens with a very large log-perplexity. This is not necessarily problematic, as the original LLM already generates these very large log-perplexities.

\begin{figure}[ht]
\vskip 0.2in
\begin{center}
\centerline{\includegraphics[width=0.4\columnwidth]{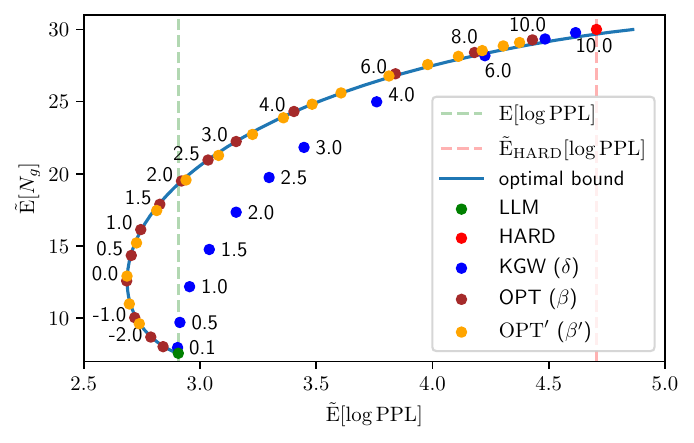}}
\caption{same as Figure \ref{fig:Ng_logPPL}, but now including the $\mathrm{OPT}^\prime$ watermark. Note that OPT and $\mathrm{OPT}^\prime$ hardly differ in terms of test-text trade-off, when text quality is defined in terms of expected log-perplexity. Error bars (vertical and horizontal) are never larger than the marker sizes.}
\label{fig:Ng_logPPL_prime}
\end{center}
\vskip -0.2in
\end{figure}

\begin{figure}[ht]
\vskip 0.2in
\begin{center}
\centerline{\includegraphics[width=0.4\columnwidth]{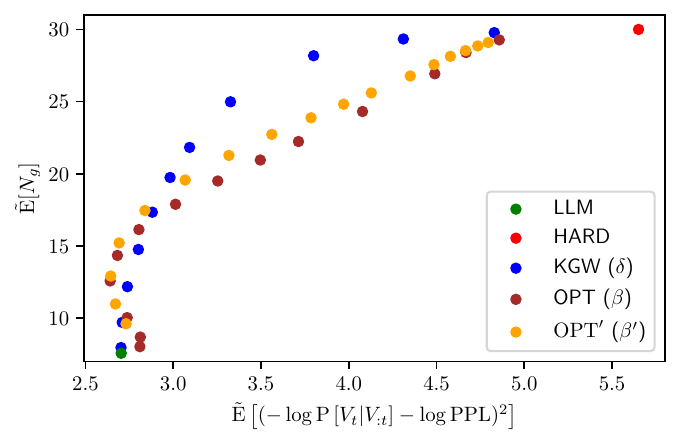}}
\caption{Test quality, measured as the expected number of green-list tokens, versus between-token variance of the log-perplexity, is shown for different watermarks. For completeness, the original language model without watermark is included (LLM). Error bars (vertical and horizontal) are never larger than the marker sizes. Note that, as expected, $\mathrm{OPT}^\prime$ outperforms OPT, albeit marginally. Also note that the KGW watermark outperforms the optimized watermarks. This is not in contradiction with our method, as we did not optimize for this trade-off.}
\label{fig:Ng_variance_prime}
\end{center}
\vskip -0.2in
\end{figure}

\begin{figure}[ht]
\vskip 0.2in
\begin{center}
\centerline{\includegraphics[width=0.4\columnwidth]{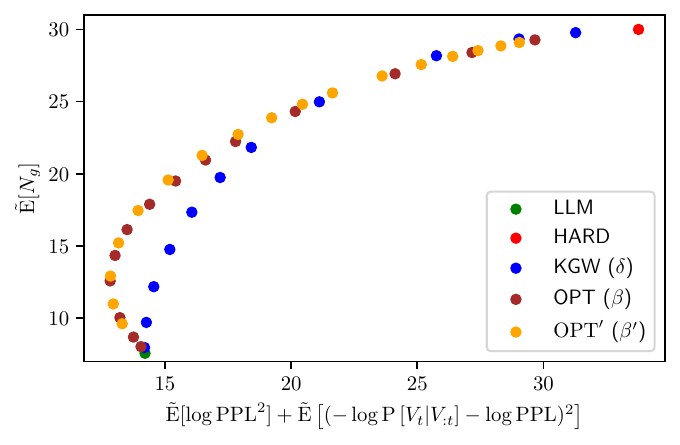}}
\caption{Test quality, measured as the expected number of green-list tokens, versus expected log-perplexity squared plus between-token variance of the log-perplexity, is shown for different watermarks. For completeness, the original language model without watermark is included (LLM). Error bars (vertical and horizontal) are never larger than the marker sizes. Note that, as expected, $\mathrm{OPT}^\prime$ outperforms both OPT and KGW, as this trade-off is the optimization objective for which $\mathrm{OPT}^\prime$ is optimized.}
\label{fig:Ng_logPPLsq_prime}
\end{center}
\vskip -0.2in
\end{figure}

Figures \ref{fig:Ng_logPPL_prime}, \ref{fig:Ng_variance_prime} and \ref{fig:Ng_logPPLsq_prime} are the same experiments as described in Section \ref{sec:experiments_test_text_trade_off}, but now including the $\text{OPT}'$ watermark. They were conducted with the OPT-350m model.

\end{document}